\documentclass[%
 reprint,
 amsmath,amssymb,
 aps
]{revtex4-1}

\usepackage{braket}
\usepackage{graphicx}
\usepackage{dcolumn}
\usepackage{bm}
\usepackage{xcolor}

\begin{document}

\title{Role of Spin-Orbit Coupling in High-order Harmonic Generation Revealed by Super-Cycle Rydberg Trajectories}

\author{N. Mayer$^{1}$,
S. Beaulieu$^{2}$,
A. Jimenez-Galan$^{1}$,
S. Patchkovskii$^{1}$,
O. Kornilov$^{1}$,
D. Descamps$^{2}$,
S. Petit$^{2}$,
O. Smirnova$^{1}$,
Y. Mairesse$^{2}$,
M. Y. Ivanov$^{1,3,4}$} 

\affiliation{
$^1$Max-Born-Institute, Max-Born str. 2A, 12489 Berlin, Germany \\
$^2$Universit\'e de Bordeaux - CNRS - CEA, CELIA, UMR5107, F33405 Talence, France\\
$^3$Department of Physics, Humboldt University, Newtonstr. 15, D-12489 Berlin, Germany\\
$^4$Blackett Laboratory, Imperial College London, SW7 2AZ London, United Kingdom.\\
}

\date{\today}

\begin{abstract}
High-harmonic generation is typically thought of as a sub-laser-cycle process, with the electron's excursion in the continuum lasting a fraction of the optical cycle. However, it was recently suggested that long-lived Rydberg states can play a particularly important role in atoms driven by the combination of the counter-rotating circularly polarized fundamental light field and its second harmonic. Here we report direct experimental evidence of long and stable Rydberg trajectories contributing to high-harmonic generation. We confirm their effect on the harmonic emission via Time-Dependent Schr{\"o}dinger Equation simulations and track their dynamics inside the laser pulse using the spin-orbit evolution in the ionic core, utilizing the spin-orbit Larmor clock. Our observations contrast sharply with the general view that long-lived Rydberg orbits should generate negligible contribution to the macroscopic far-field high harmonic response of the medium. Indeed, we show how and why radiation from such states can lead to well collimated macroscopic signal in the far field.

\end{abstract}

\maketitle


The interaction of the electron's spin with its orbital motion, the spin-orbit coupling (SOC), is important in atoms, molecules, and solids. This coupling leads to the spin-orbit energy splitting of states (energy $\Delta_{SO}$). Excitation of the spin-orbit-split states upon ionization can either induce ultrafast hole dynamics in the ion
\cite{Goulielmakis10,Kuebel2019,Kobayashi20}, or generate entanglement between the ion and the photoelectron \cite{smirnova2010spin}, depending on the degree of coherence created upon ionization \cite{Goulielmakis10,smirnova2010spin,Pabst2016,ruberti2019onset}. The latter can be fully controlled with modern attosecond technologies \cite{vrakking2021control}.

Yet, the role of SOC interaction in High-Harmonic Generation (HHG), the process lying at the core of attosecond technologies, and the opportunities it presents for time-resolved spectroscopy of systems in strong laser fields, have largely remained unexplored. Indeed, the time-delay $\tau=t_r-t_i$ between field-induced electron tunnelling ($t_i$) and its radiative recombination ($t_r$) with the ionic core, emerging in the three-step picture of HHG \cite{Lewenstein:1994aa,corkum1993plasma},  generally lasts only a fraction of the laser cycle, $\tau\sim 1/\omega$, too short \cite{Pabst:2014aa} for resolving  spin-orbit dynamics in typical media used for HHG, such as argon or neon in which $\tau\Delta_{SO}\ll 1$, unless very long-wavelength drivers are used \cite{Pabst:2014aa}. The microscopic contribution of long electron trajectories with $\tau\sim 1/\Delta_{SO}\gg 2\pi/\omega$, compatible with resolving the spin-orbit dynamics, is believed to be weak in the macroscopic HHG signal due to poor phase-matching and extremely large spatio-spectral spreading of such long trajectories.
Here we use the built-in Larmor clock offered by spin-orbit interaction \cite{kaushal2015spin} to identify both the surprisingly prominent contribution of long-lived Rydberg orbits to the far-field HHG signal and the physics responsible for their appearance. Our work also confirms the results proposed in \cite{Jimenez:2017aa} that Rydberg states lead to the generation of symmetry-forbidden harmonics. 

We study HHG in argon, where coherent excitation of the spin-orbit-split doublet of the $P_{3/2}$ and $P_{1/2}$ core states ($\Delta_{SO}=0.177$ eV) leads to the precession of the spin of the hole  with a period $T_{SO}=23.5$ fs. If the tunneled electron returns to the core when the hole spin is flipped, recombination is quantum-mechanically prohibited. As a result, the photorecombination amplitude is temporally modulated by the factor $1+\exp(-i\Delta_{SO}\tau)$. If $\tau$ spans at least $\tau\simeq T_{SO}$, this temporal modulation of the emission will spectrally split the harmonic lines by $\Delta_{SO}$. In contrast, if the recombination events contributing to HHG occur within a fraction of the laser cycle with $\omega\gg \Delta_{SO}$, no spin-orbit splitting will be seen \cite{Pabst:2014aa}. Experimentally, we observe well-collimated high harmonic emission split by $\Delta_{SO}$ in Ar driven by circularly polarized $800$ nm fundamental field and its counter-rotating second harmonic. Our results show that this splitting can only occur due to long-lived electron orbits lasting at least $T_{SO}\simeq24$ fs. 

Our theoretical analysis explains how multiphoton excitation of Rydberg states leads to the emergence of spin-orbit splitting in harmonic spectra. Moreover, it allows us to show how such multiphoton Freeman resonances \cite{Freeman:1987aa,Gibson:1994aa} shape the harmonic emission in time and space, and how their contribution emerges in the far-field.

\begin{figure*}
\begin{center}
\includegraphics[width=15cm, keepaspectratio=true]{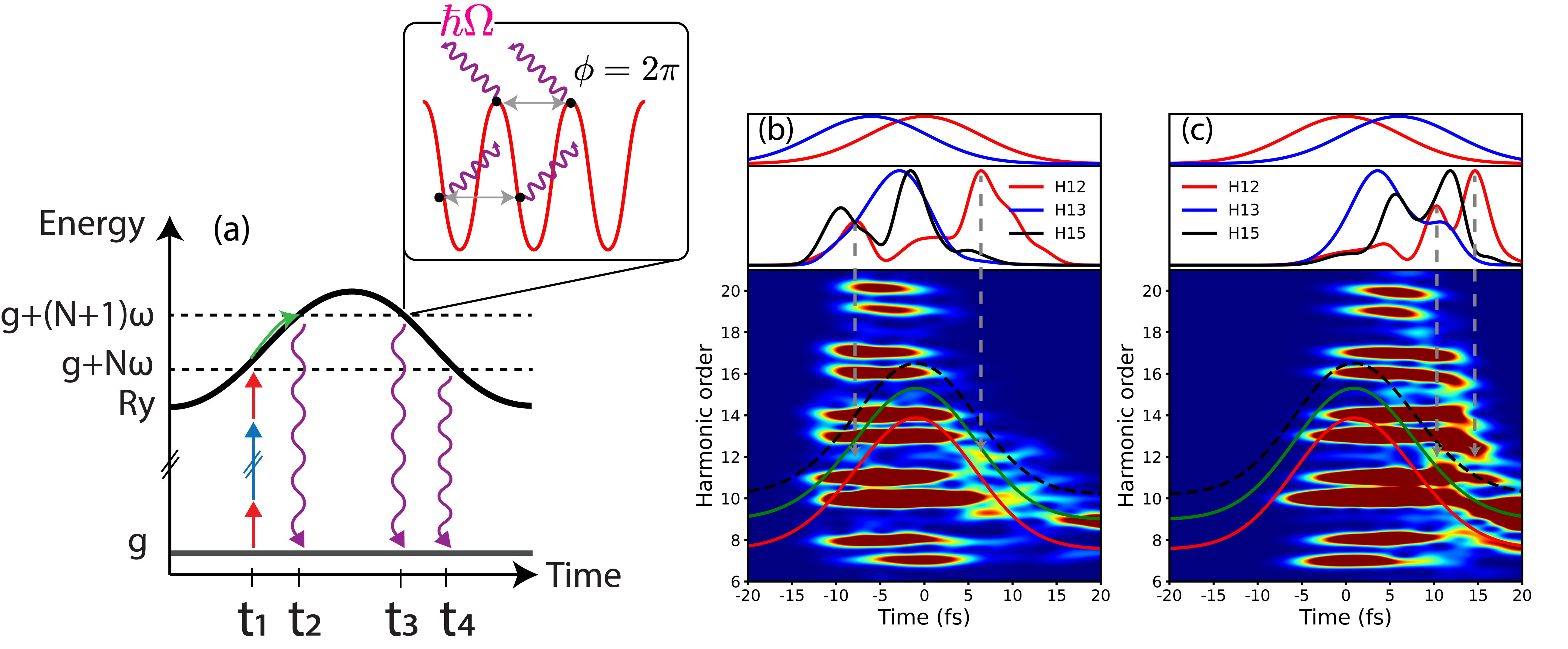}
\caption{\textbf{The role of Rydberg state in multi-cycle bicircular HHG.} \textbf{(a)} Resonant multiphoton transitions to Stark-shifted Rydberg states at the rising ($t_1$) or falling edge ($t_4$) or near the peak of the pulse ($t_2$, $t_3$). Cycle-to-cycle interference (inset) leads to resonant emission optimized at integer multiples of the fundamental frequency $m\omega$. \textbf{(b,c)}: Calculated Gabor spectrogram (Gaussian window of $\text{FWHM}=2\pi/\omega$) of the harmonic emission obtained by solving the TDSE in argon. The time-dependent energy of the ponderomotive-shifted ionization potential (dashed black line), $2p^53d$ state (green solid line) and $2p^54s$ state (red solid line) are also shown. Inset (\textbf{(b,c)}): time-dependent emission of selected harmonics, i.e. symmetry-allowed H13 and forbidden H12 and H15. In \textbf{(b)} the $2\omega$ field precedes the $\omega$ field ($\tau=6$ fs),  vice versa in \textbf{(c)} ($\tau=- 6$ fs). Both fields have an ellipticity of $\epsilon=0.9$.}
\label{Fig1}
\end{center}
\end{figure*}

Fig.~\ref{Fig1}(a) shows how resonantly-populated Rydberg states participate in the HHG process
at the single atom level~\cite{Beaulieu:2016aa,camp18}. They are populated through light-induced Freeman resonances \cite{Freeman:1987aa,Gibson:1994aa}, the frequency-domain counterpart of the time-domain frustrated tunnelling \cite{Zimmermann:2017aa}. Resonances can arise on the rising and falling edges of the pulse, or near its peak (times $t_{i}$ of Fig. \ref{Fig1}(a)), when

\begin{equation}
E_n+U_p(t)-E_g=n_\omega \omega+n_{2\omega}2\omega
\end{equation}

where $n_{r\omega}$ is the number of absorbed photons from the $r\omega$ field, $E_n$ is the field-free energy of the Rydberg state, and the AC Stark shift is approximated by the ponderomotive energy $U_p(t)=U^\omega_p(t)+U^{2\omega}_p(t)$, with  $U_p^{r\omega}(t)=E^2_{r\omega}(t)/2(r\omega)^2$. 

Concomitant with resonant excitation at times $t_1$, resonant emission from Rydberg states can occur. The resonant emission is enhanced by constructive cycle-to-cycle interference.  Indeed, for a Rydberg orbit, the relative phase $\phi$ accumulated between the two successive emission events separated by the optical cycle $T_{cyc}$ is $\phi=(E_n+U_p-E_g)T_{cyc}$. Their constructive interference requires $\phi=(E_n+U_p-E_g)T_{cyc}=2\pi m$, i.e. $E_n+U_p-E_g=m\omega$, optimizing emission from the Rydberg states at integer harmonic orders and on resonance with the ground state.

Importantly, Rydberg excitations can lead to the emission of symmetry-forbidden (e.g. parity forbidden) harmonics, which in the specific case of bicircular $(\omega,2\omega)$ fields are harmonics with order 3N \cite{Jimenez:2017aa}. Let excitation on the rising edge (time $t_1$ in Fig. 1(a)) populate a Rydberg state that can recombine via a dipole transition to the ground state. After excitation, the excited Rydberg state is shifted up in energy (green line in Fig. 1(a)). At a later time ($t_2$ or $t_3$ in Fig. 1(a)), the Rydberg state again enters a multi-photon resonance with the ground state. Even if the resonant energy formally corresponds to symmetry-forbidden harmonic lines, radiative recombination can occur without violating conservation rules since the parity and the angular momentum of the Rydberg state is conserved by the ponderomotive shift. Forbidden harmonic generation is thus an evidence of a super-cycle Rydberg trajectory that starts at a resonant time $t_1$ and recombines later at  a resonant time $t_2$.
The combination of the cycle-to-cycle interference and the ponderomotive shift of the Rydberg state  thus effectively act as a temporal gate on the dipole moment $\mathbf{d}(t)$, enhancing the harmonic emission  near the resonant times $t_i$, at both allowed and forbidden frequencies.

To support this picture we have performed TDSE simulations in Ar using the code of Patchkovskii \textit{et al.} \cite{Patchkovskii:2016aa}. We use a bicircular field composed  of two Gaussian pulses centered at 800 nm and 400 nm,  both of 15 fs FWHM duration, with peak intensities of  $I_\omega=I_{2\omega}=7\cdot10^{13}$ W/cm$^2$. To account for experimental polarization state artifacts, we consider pulses with ellipticity $\epsilon_{\omega}=\epsilon_{2\omega}=0.9$ (calculations for circularly polarized light $\epsilon_{\omega}=\epsilon_{2\omega}=1.0$ show no qualitative differences.) At the peak of the pulse, the total ponderomotive shift is $U_p\simeq7\omega$.  Figures~\ref{Fig1}b,c) show the Gabor spectrogram of the  dipole acceleration $\ddot{\mathbf{d}}(t)$ obtained 
using a Gaussian window function with FWHM of one optical cycle $T_{cyc}=2\pi/\omega$. As shown in Ref.~\cite{Jimenez:2017aa}, harmonic generation from Rydberg states is enhanced in a bicircular pulse  when the pulses are not overlapping and when the $2\omega$ pulse precedes the $\omega$ one. We thus construct a time-frequency spectrogram (the Gabor spectrogram) of the harmonic emission for two relative delays: $\tau=6$ fs in figure~\ref{Fig1} b) where $2\omega$ precedes $\omega$, and $\tau=-6$ fs in figure~\ref{Fig1} c), when $\omega$ precedes $2\omega$ one.

The dashed black line shows the ponderomotive shift of the ionization 
potential, while the solid red and green lines show the ponderomotive shift of the Rydberg states 2p$^5$4s  and 2p$^5$3d, respectively. The top panel shows the time-dependent emission for  H12, H13 and H15, obtained by integrating the Gabor spectrogram around the corresponding energies, in the $\pm 0.25\omega$ range. For both forbidden harmonics H12 and H15 the contribution of Rydberg states is more visible than for the  allowed H13, where short trajectories dominate the emission. Due to imperfect circularity of the pulses, absorption of counter-rotating components from either field can now lead to 3N harmonic orders. Nonetheless, resonant enhancement of these orders is clearly observed in both Fig. 1(b) and 1(c).

\begin{figure*}
\begin{center}
\includegraphics[width=14cm, keepaspectratio=true]{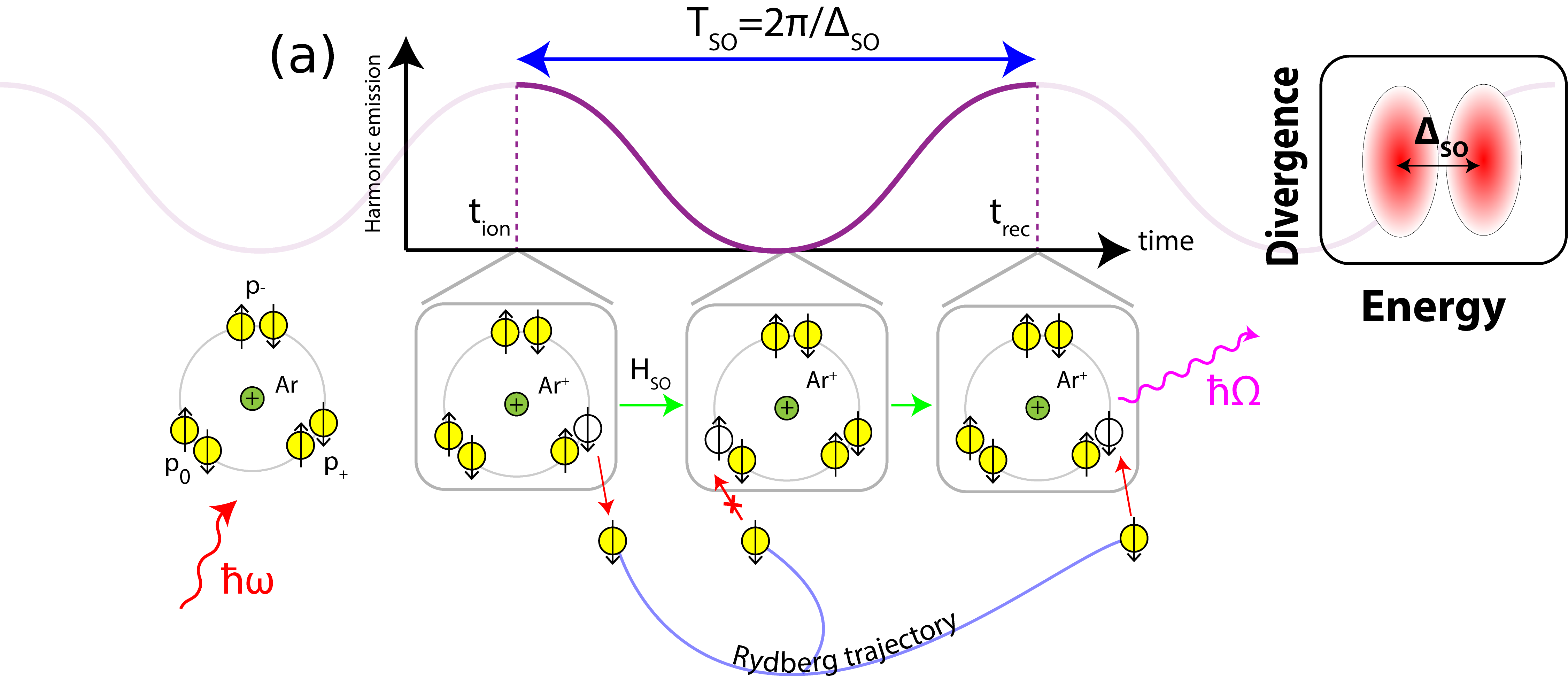}
\caption{\textbf{Spin-orbit temporal modulation of the emission leading to spectral splitting of the harmonic lines.} \textbf{(a)} Spin-orbit Larmor clock in HHG. Radiative electron-hole recombination is temporally modulated due to the precession of the spin of the hole, leading to a spectral splitting of the harmonic lines.}
\label{Fig2}
\end{center}
\end{figure*}

When the second harmonic precedes the fundamental pulse (Fig. \ref{Fig1} b)), enhanced emission of both H12 and H15 is observed near -7 fs. At this time, the Rydberg states just below the ionization threshold, and the 2p$^5$4s state enter resonance with the two adjacent harmonics. In agreement with figure \ref{Fig1} a), we interpret the enhancement as a signature of Rydberg states excited by a multiphoton resonance that can later recombine at a "forbidden" harmonic. Similarly at around 0 fs, enhancement of H12 and H15 is due to the multiphoton resonance of the 2p$^5$3d state at the H15 energy. On the falling edge ($\sim 7$ fs) the H12 is enhanced again due to the resonance with the 2p$^5$3d state. 

When the fundamental pulse precedes its second harmonic (Fig. \ref{Fig1} c)) the shape of the temporal gate changes drastically: resonant enhancements are seen mainly on the falling edge. This asymmetry can be interpreted via Frustrated Tunneling Ionization (FTI) \cite{yudin2001physics, Nubbemeyer:2008aa}. Efficient trapping in a Rydberg orbit requires that the lateral velocity shift $A_\omega=E_\omega/\omega$, imparted on the tunneled electron by the red field, is compensated by the blue field, whose vector potential is two times weaker for equal field strengths. 
Thus, at positive time delays (when $2\omega$ precedes $\omega$) the blue field is stronger on the rising edge of the total pulse and FTI occurs at the rising edge. For negative delays, 
the compensation of the lateral shift needed for efficient FTI occurs only on the pulse falling edge, consistent with resonant HHG enhancement then. Importantly, we see that the time the trapped Rydberg population spends inside the pulse is longer when the $2\omega$ field precedes the $\omega$ one (positive delays).

\begin{figure*}
\begin{center}
\includegraphics[width=14cm, keepaspectratio=true]{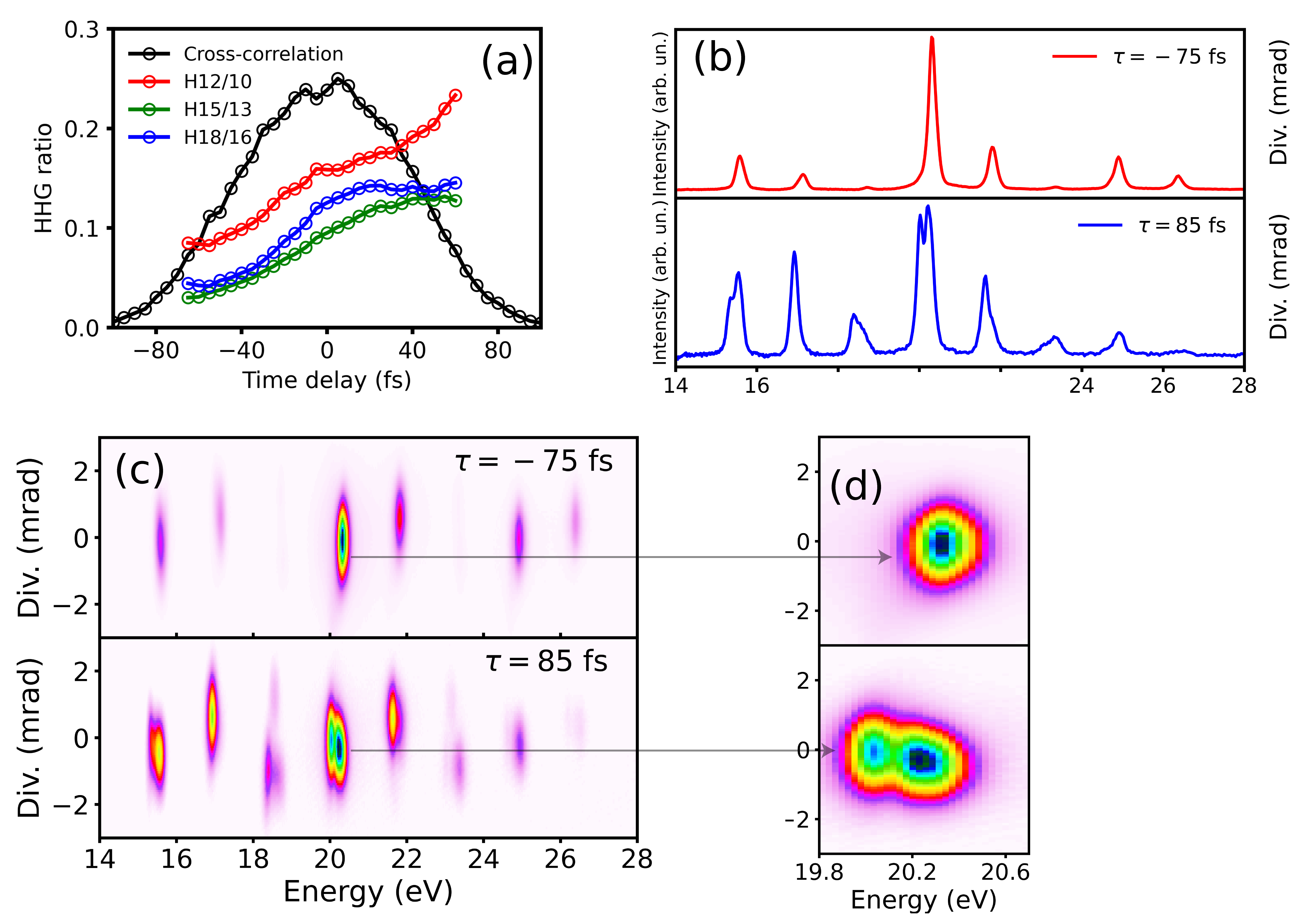}
\caption{\textbf{Experimental results} \textbf{(a)} Measured delay-dependent total harmonic yield (black) and ratio between 3N and 3N-2 harmonics: H12/H10 (red), H15/H13 (green), H18/H16 (blue). \textbf{(b,c)} Far-field HHG spectrum with $\omega$ field preceding ($\tau$ = -75 fs) or following ($\tau$ = 85 fs) the $2\omega$ field. For $\tau=85$ fs, enhanced generation of forbidden harmonics is observed. The spin-orbit splitting of the harmonic lines is clearly seen in the spatially resolved harmonic spectrum in \textbf{(c)}. \textbf{(d)} Spatially- and energy-resolved profile of H13 at negative and positive delays.}
\label{Fig3}
\end{center}
\end{figure*}

We now take advantage of this temporal gating of the HHG emission to resolve the spin-orbit hole dynamics in the core. The well-defined start of the spin-orbit clock is provided by the light-induced multiphoton resonance: the time window during which the ponderomotive shift brings a Rydberg state into a multiphoton resonance with the ground state is much shorter than the pulse duration. Upon excitation, the emission from the Rydberg state is modulated in time with the spin-orbit period, leading to the splitting of the harmonic lines by $\Delta_{SO}$ (see Fig. \ref{Fig2}). The appearance of the splitting is the evidence of the long-lived Rydberg orbits. As noted above, the time spent by the trapped Rydberg orbits during the pulse is substantially longer when 2$\omega$ precedes the $\omega$: this is where the spin-orbit splitting is expected to appear. 

We performed experiments in argon, with $\Delta_{SO}=0.177$ eV corresponding to $T_{SO}=23.5$ fs, using the CELIA Ti:Sapphire Aurore laser system, delivering  up to 7 mJ energy per pulse centered at 800 nm, at 1 kHz repetition rate, with pulse duration of 25 fs. To favor the population of Rydberg states, slightly chirped pulses were used for the experiment with a duration of about 50 fs. The pulses were split in a 50$\%$-50$\%$ Mach-Zehnder interferometer. One of the two arms was frequency-doubled in a 200 $\mu$m thick type I BBO crystal. In each arm, we placed a broadband zero-order quarter wave-plate to control the polarization states of the 800 nm  and the 400 nm pulses independently, generating close to circular, counter-rotating fields. The pulses were focused onto a $\sim$ 250 $\mu m$ thick effusive Ar gas jet. The XUV radiation was analyzed by a flat-field XUV spectrometer, consisting of  a 1200 grooves/mm cylindrical grating (Shimadzu) and a set of dual microchannel plates coupled to a fast P46 phosphor screen (Hamamatsu) enabling single shot measurements. A 12-bit cooled CCD camera (PCO) recorded the spatially-resolved harmonic spectra  for different $\omega-2\omega$ delays.

Fig. ~\ref{Fig3}(a) shows the ratio between selected forbidden and allowed harmonic orders as a function of the relative time delay, where for positive time delays the $2\omega$ field precedes the $\omega$ one. The delay dependence of the forbidden harmonics confirms the behaviour predicted in \cite{Jimenez:2017aa} and our TDSE simulations: the participation of Rydberg states in HHG is optimized when $2\omega$ precedes $\omega$. Figure 3(b) shows the harmonic spectrum at positive and negative delays. At positive delays
we see strong enhancement of forbidden 3N orders due to Rydberg states. Moreover, we observe spectral splitting of each harmonic consistent with the spin-orbit energy of argon in the HHG spectrum. The splitting is clearly visible in the spatially resolved far-field spectrum in Figure 3c), where we are able to spatially separate different photon channels using a small crossing angle between the two beams \cite{Hickstein:2015aa}. The thin gas jet used in the experiment excludes longitudinal phase matching and propagation effects. Therefore we conclude that, by trapping the Rydberg electron for $\sim 24$ fs at intensities of $I\sim 2\times 10^{14}$W/cm$^2$, we resolve the spin precession of the electron hole in the core. The absence of the spectral splitting at negative time delays is consistent with  the discussion above: when $\omega$ precedes  $2\omega$, FTI occurs at the falling edge of the pulse, leaving insufficient time to resolve the spin-orbit-induced modulation.

Our experiment shows the contribution of very long trajectories in the far-field, lasting ten (or more) cycles. Yet, it is known that electron trajectories spending just about one laser cycle (or longer) in the 
continuum give little contribution to the macroscopic signal \cite{Gaarde:2002aa}.  The reason for the suppression of the long trajectories' contribution  is the strong dependence of the harmonic phase 
$\phi_N$  on the laser intensity $I$, $\phi_N\simeq\alpha_N I$, where $\alpha_N$ is roughly proportional to electron's excursion time. Large $\alpha_N$ corresponds to a strongly curved temporal and spatial phase, leading to a large spectral broadening and spatial divergence \cite{Salieres:1995aa,catoire2016}. How can trajectories lasting for about ten laser cycles contribute to the far-field HHG spectrum?  

\begin{figure*}
\begin{center}
\includegraphics[width=10cm, keepaspectratio=true]{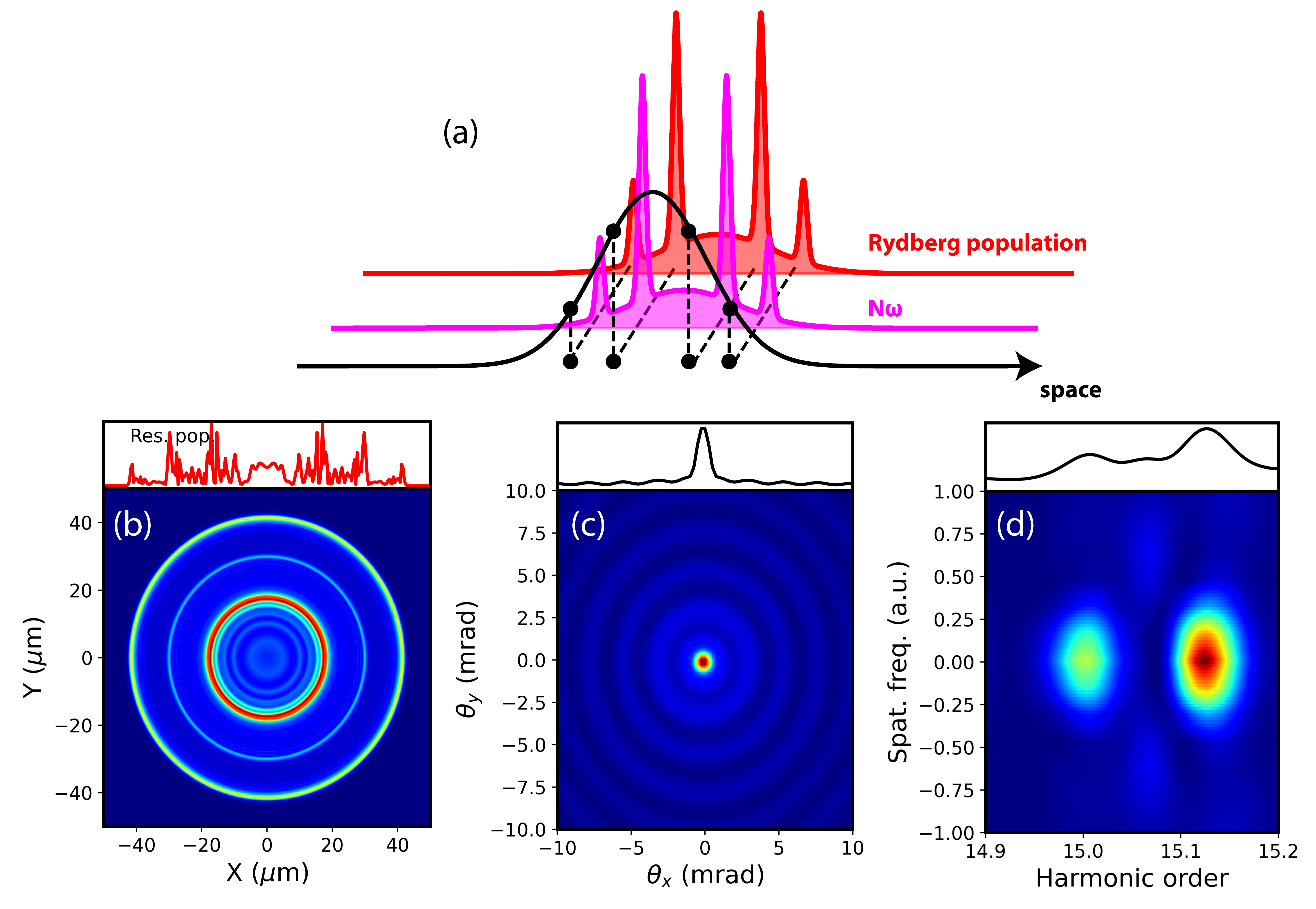}
\caption{\textbf{The role of spatio-temporal gating on the generation of collimated and spin-orbit split harmonic emission.} \textbf{(a)} The role of Rydberg states in HHG in the spatial domain. Resonant multiphoton transitions to Stark-shifted Rydberg states lead to resonant enhancements at the positions $\mathbf{x}_i$, defining onion shells where the harmonic phase is large but constant. \textbf{(b,c,d)} Results from the two-level system  model. \textbf{(b,c)} show the near- and far-field profiles of H15 respectively, while \textbf{(d)} shows the spin-orbit splitting of the far-field harmonic spectrum around H15. The top panel in \text{(b)} shows the spatial distribution of the residual population in the excited state at the end of the pulse. The top panel in \text{(c)} shows the 1D spatial distribution in the far-field of H15, while the top panel in \textbf{(d)} shows the frequency-resolved, spatially integrated far-field harmonic spectrum when spin-orbit modulations are included.}
\label{Fig4}
\end{center}
\end{figure*}

Our explanation is associated with the resonant nature of the process, which selects a specific spatial region in the laser focus, where the resonance(s) with Stark-shited Rydberg state(s) is achieved.  The intensity of the driving field is approximately constant along this ring. There,  the resonant contribution to the harmonic emission is maximally enhanced, and, moreover, the phase of the emission is fixed, leading to a collimated far-field pattern. The same spatial region provides the dominant contribution to Freeman resonances in photoelectron spectra \cite{Gibson:1994aa,Agostini:1993aa}. Thus, the light-induced multiphoton resonances with Rydberg states provide an addition spatial gating to the harmonic emission, see Figure \ref{Fig4}a), complementary to the time-domain gating explained earlier.

To confirm this picture, we model  high-harmonic generation from an excited state using the two-level system (TLS) of Ref. \cite{Bloch:2019aa}. The TLS is driven by the field $E(x,t)=g(x)f(t)E_0\sin\omega t$, with $E_0=0.08$ a.u., $\omega=0.056954$ a.u. ($\lambda=800$ nm). Here $g(x)$ describes a Gaussian spatial intensity with width of $\sigma_x=40\mu$m, and $f(t)$ is a Gaussian envelope with FWHM of 2000 a.u. ($\simeq$50 fs). The transition energy is $\Delta=0.56$ a.u. and the dipole coupling between the ground and the excited states is set to $d_{10}=\sqrt{\Delta}/2\omega=6.57$ a.u. to mimic the ponderomotive shift of the Rydberg state \cite{Bloch:2019aa} and the associated intensity-dependent phase accumulated by the excited state. The time-dependent dipole moment at the position $x$ is  
$d(x,t)=d_{10}\text{Re}[c_0(t)c^*_1(t)]$, where $c_{k}(t)$ are found by solving the coupled equations
\begin{eqnarray}\text{i}\dot{c}_0&=&-\frac{\Delta}{2}c_0(t)-d_{10}E(x,t)c_1(t)\nonumber\\
\text{i}\dot{c}_1&=&\frac{\Delta}{2}c_1(t)-d_{10}E(x,t)c_0(t)-\text{i}\gamma|E(x,t)|^2,\end{eqnarray}
with initial conditions $c_0(0)=1,\,c_1(0)=1$. The decay term with $\gamma=0.3$ describes ionization of the excited state. To mimic the effect of the spin-orbit interaction, we multiply the dipole moment by the oscillating term $\left(1+\exp(-i\Delta_{SO}t)\right)$, where $\Delta_{SO}=0.007$ a.u. is the spin-orbit energy. The far-field harmonic spectrum  is computed via 
spatial and temporal Fourier transforms.

The near-field profile of H15 (Fig. \ref{Fig4}b)) clearly shows resonant enhancements when the multiphoton resonance is reached at the peak of the field. The residual population of the excited state, shown above the near-field profile in red, peaks at the same positions.  In the far-field (Figure \ref{Fig3}c)), we observe well-collimated harmonic emission. The splitting of the harmonic due to the spin-orbit modulation is seen in Figure \ref{Fig4}d. The low intensity of the outside rings in Figure \ref{Fig4}c prevents their observation in the experiment.

Our work sheds light on the role of Rydberg states in the HHG process. The influence of excited states in HHG has been the subject of many experimental \cite{Chini14, Beaulieu:2016aa, Beaulieu17a, Bloch:2019aa, Yun18} and theoretical investigations \cite{Taieb:2003aa, Bian10, Camp:2015aa}. On very short time-scales, the excitation of Rydberg states followed by ionization and recombination onto the ground state was shown to lead to a delayed (by few femtoseconds) emission of an attosecond pulse train \cite{Beaulieu:2016aa}. On longer timescales, the population of Rydberg states can induces hyper-Raman lines \cite{Millack93,Bloch:2019aa} if a dressing laser field is added. Moreover, on timescales much longer than the laser pulse duration, the Rydberg wavepacket generates coherent XUV free-induction decay (FID) \cite{Bengtsson17,Beaulieu17a}. Here we work in an intermediate range, where we demonstrate the coherent emission of XUV radiation through strong-field driven electron trajectories lasting several tens of femtoseconds. The identified mechanism provides a link between the Freeman resonances and the HHG process \cite{Freeman:1987aa, Taieb:2003aa, Taieb:2003ab}. Our work also shows that using the spin-orbit Larmor clock allows one to monitor  the presence of Rydberg orbits, living at least $\sim 24$ fs, inside a laser pulse with intensities of 200 TW/cm$^2$. It also opens perspective of probing the dynamics of electronic wavepackets on tens of femtoseconds, by extending high-order harmonic spectroscopy beyond the optical cycle. This would be of particular interest to reveal the role of intervalley scattering, as well as coherent exciton dynamics and its decoherence in HHG from photoexcited two-dimensional materials \cite{Rustagi19,Christiansen19,dong21}.

N. M. and M. I. acknowledge support from DFG QUTIF:
IV 152/6-2. Á. J-G. acknowledges support from the H2020
Marie Skłodowska-Curie Actions (101028938). This
project has received funding from the European
Research Council (ERC) under the European Unions
Horizon 2020 research and innovation Program
No. 682978 - EXCITERS, and from the French National
Research Agency through ANR-14-CE32-0014 MISFITS.

\bibliography{biblio_new}
\end{document}